\documentclass[]{spie}  %>>> use for US letter paper
\usepackage{amsmath,amsfonts,amssymb}
\usepackage{graphicx}
\usepackage[colorlinks=true, allcolors=blue]{hyperref} 
\usepackage{multirow}

\title{The Balloon‑borne VLBI Experiment (BVEX) Radio Telescope and Position Tracking System}

\author[a]{Felix M. Thiel}
\author[a]{Mayukh Bagchi}
\author[a]{Laura M. Fissel}
\author[a]{Aarchi Shah}
\author[b,c]{Lindy Blackburn}
\author[d]{Emily Butler}
\author[e]{Rafael Costa}
\author[f]{Thomas Emo}
\author[g]{Vincent L. Fish}
\author[h,i]{Daryl Haggard}
\author[b,c]{Michael D. Johnson}
\author[j]{Jessica Lo}
\author[a]{Maggie Oxford}
\author[b,c]{Dominic W. Pesce}
\author[g]{Ganesh Rajagopalan}
\author[k]{Javier L. Romualdez}
\author[l]{Adrian K. Sinclair}
\author[m]{Bonnie Slocombe}
\author[a]{Stephanie St-Jean}
\author[e]{Terry Yang}

\affil[a]{Queen's University, Department of Physics, Engineering Physics and Astronomy, 64 Bader Lane, Kingston, ON K7L 3N6, Canada}
\affil[b]{Center for Astrophysics | Harvard \& Smithsonian, 60 Garden Street, Cambridge, MA 02138, USA}
\affil[c]{Black Hole Initiative, Harvard University, 20 Garden Street, Cambridge, MA 02138, USA}
\affil[d]{Dalhousie University,Department of Physics and Atmospheric Science,1453 Lord Dalhousie Drive, Halifax, NS  B3H 4R2, Canada}
\affil[e]{Queen's University, Smith School of Engineering, 45 Union Street, Kingston, ON K7L 2N8, Canada}
\affil[f]{Queen's University, Department of Mathematics and Statistics, 48 University Avenue, Kingston, ON K7L 3N6, Canada}
\affil[g]{MIT Haystack Observatory, 99 Millstone Road, Westford, MA 01886, USA}
\affil[h]{Trottier Space Institute, McGill University, 3550 rue University, QC H3A 2A7, Canada}
\affil[i]{McGill University, Department of Physics, 3600 rue University, QC H3A 2T8, Canada}
\affil[j]{McMaster University, Department of Engineering Physics, 1280 Main Street West, Hamilton, ON L8S 4L7, Canada}
\affil[k]{StarSpec Technologies Inc., 402 Harmony Road Units 8-11, Ayr, ON N0B 1E0, Canada}
\affil[l]{Johns Hopkins University, Bloomberg Center for Physics and Astronomy, 3400 North Charles Street, Baltimore, MD 21218, USA}
\affil[m]{University of British Columbia, Department of Physics \& Astronomy, 325 - 6224 Agricultural Road, Vancouver, BC V6T 1Z1, Canada}
\authorinfo{Send correspondence to F.M.T\\E-mail: 21fmt2@queensu.ca}

\begin{document} 
\maketitle

\begin{abstract}
Very Long Baseline Interferometry (VLBI) is a technique in radio astronomy that vastly increases the diffraction limited angular resolution of radio observations by correlating simultaneous observations of radio telescopes separated by thousands of kilometers. Ground-based high-frequency VLBI is limited in baseline length by the diameter of the Earth and in sensitivity by molecular absorption of the Earth’s atmosphere. One of the ways to improve VLBI is with balloon telescopes, which are much cheaper than space VLBI telescopes. In addition balloon-VLBI stations could also improve image quality due to increased uv-coverage. As a first step to achieving VLBI from a balloon-borne platform we present the Balloon-borne VLBI Experiment (BVEX).This is a novel, first-generation balloon-borne VLBI station observing at K-band (22 GHz) which launched as part of the Canadian Space Agency (CSA) STRATOS campaign from Timmins, Ontario, Canada in August 2025, with the aim to demonstrate VLBI from a balloon-borne platform by correlating with a large ground based telescope. We discuss the design of the 36 inch (91 cm) radio telescope and receiver with a 2 GHz bandwidth built from commercial-off-the-shelf (COTS) components as well as the telescope pointing system and weld-free mounting structure capable of interfacing with the CSA’s CARMENCITA gondola. We also present a novel high precision position tracking system built from COTS components giving $<$1 mm precision on 1 s time scales which is required for achieving phase tracking at 22 GHz.  We further show on-the-ground testing results from the 2025 Timmins campaign. The 2025 flight from Timmins was not successful due to a leak in the balloon preventing the payload from reaching its target altitude where accurate gondola pointing is possible.  We therefore wrap up by presenting plans to upgrade the telescope for a potential next flight opportunity from Palmas, Tocantins, Brazil in 2027.
\end{abstract}

\section{INTRODUCTION}
\label{sec:intro}  % \label{} allows reference to this section

Very Long Baseline Interferometry (VLBI) is a technique in radio astronomy that combines simultaneous observations from radio telescopes across the Earth effectively giving the resolution of an Earth-sized telescope given by 
\begin{equation}
    \theta \approx \frac{\lambda}{b_{max}}
\end{equation}
where $\lambda$ is the observing wavelength and $b_{max}$ is the largest separation between telescopes (baseline). This technique has been leveraged by the Event Horizon Telescope (EHT) to image the shadows of the super-massive black holes at the center of M87 \cite{EHT2019M87I} and the Milky Way \cite{EHT2022SGRAI} at a frequency of 230 GHz (1.3 mm) as well as the Global Millimetre VLBI Array (GMVA) to image the jet in the vicinity of M87* \cite{Lu2023} at a frequency of 86 GHz (3 mm). Ground-based high-frequency VLBI in its current form is limited in two ways. Firstly, it is limited in resolution due to the size of the Earth and secondly in dynamic range (i.e. the ability to resolve objects of different sizes) by the fact that not all ground-based sites on Earth are suitable for high-frequency observing due to absorption of cosmic signals by water vapor in the Earth's atmosphere. In recent years there has been a rising interest in space-VLBI which has the ability to increase baseline lengths beyond the limit imposed by the size of the Earth through missions like the Black Hole Explorer (BHEX)\cite{Johnson2024}. A cheaper, and much more affordable option is to build a balloon-borne VLBI station that complements a ground-based telescope array and operates at an altitude of about 35 km above more than 99.5\% of the Earth's atmosphere. While this does not increase the baseline length it gives access to observing locations that are inaccessible to ground-based astronomy. This not only improves the dynamic range of these observations but also enables higher frequency observations than is currently possible from the ground resulting in a resolution increase. 

In this paper, we will be presenting designs for the radio telescope and position tracking system of the balloon-borne VLBI experiment (BVEX) a 22 GHz prototype mission with the aim to demonstrate VLBI between a balloon-borne telescope and a ground-based telescope through the detection of fringes, while observing bright radio galaxies. A telescope like this has already been built and successfully ground-tested in Ref.~\citenum{Doi2019} and recently launched from Taiki Aerospace Research Field in Japan on a short duration 3.5 hour balloon flight on June 14$^{\text{th}}$ 2026. While Ref.~\citenum{Doi2019} mostly employs custom components with a full dual polarization and dual side-band system, BVEX is built entirely from commercial, off-the-shelf (COTS) components and offers a compact, scalable and cheap model for balloon-borne VLBI. BVEX is funded through the Canadian Space Agency's (CSA) FAST program. The CSA partners with the Centre National d'Études Spatiales (CNES) as a launch provider and therefore BVEX can potentially have access to a wide range of launch opportunities ranging from short duration flights lasting about 12 hours up to long-duration flights lasting for around 4 days, which would allow for long timescale observations and more Fourier coverage in the final image. We will begin by discussing the radio telescope and receiver design in Sec.~\ref{sec:telescope}. We will then present an overview of the BVEX position reconstruction system in Sec.~\ref{sec:position}, we will then proceed with a discussion of ground-based tests as well as the 2025 STRATOS flight campaign from Timmins, Ontario in Sec.~\ref{sec:ground_tests} before wrapping up with a discussion of receiver upgrades for a future flight opportunity from Palmas, Tocantins, Brazil during the Summer of 2027.

\section{RADIO TELESCOPE AND RECEIVER}
\label{sec:telescope}
When designing a radio telescope for VLBI purposes the key quantity of interest is the System Equivalent Flux Density (SEFD) given by
\begin{equation}\label{eq:SEFD}
    SEFD = \frac{2kT_{sys}}{A_0}
\end{equation}
where $A_0$ is the on-axis effective collecting area of the radio dish and $T_{sys}$ is the system noise temperature. The SEFD therefore is the flux of a point source that would increase the antenna temperature by the system temperature. The noise RMS on a particular baseline is given by \begin{equation}\label{eq:sigma}
    \sigma_{rms} = \frac{1}{\eta_q}\sqrt{\frac{SEFD_1 \times SEFD_2}{2\Delta\nu\tau}}
\end{equation} 
where $SEFD_1$ and $SEFD_2$ are the SEFD's of the two telescopes on that baseline, $\Delta\nu$ is the bandwidth of the observation, $\tau$ is the coherence time on that baseline and $\eta_q$ is the quantization efficiency (0.88 for 2-bit quantization)\cite{2017isra.book.....T}. 

For the design of BVEX we expect our coherence time to be limited to 1s by both the internal crystal oscillator clock (OCXO) as well as the position tracking system (see Sec.~\ref{sec:position}). Furthermore, because BVEX will be observing at K-band (22 GHz) we will assume that our instantaneous bandwidth is limited to 512 MHz which is the maximum instantaneous bandwidth of VLBI arrays like the Very Long Baseline Array (VLBA) and the High Sensitivity Array (HSA). The other design consideration arises from the fact that Eq.~\ref{eq:sigma} is proportional to the geometric average of the SEFD's of both stations on a particular baseline. We can therefore compensate for the fact that the collecting area for BVEX will be small and the receiver will be non-cryogenic by correlating with a large collecting area ground-station. The design goal therefore was to design a compact radio telescope from commercial-off-the-shelf (COTS) components with a room-temperature receiver that would be able to detect bright ($>$10 Jy, unresolved flux) calibrators on baselines with existing ground-based K-band facilities like the VLBA (SEFD: 640 Jy), the Green Bank Telescope (GBT, SEFD: 20 Jy), the Effelsberg 100 m (SEFD: 70 Jy) as well as the Haystack 37m (SEFD: 1160 Jy)\footnote{\url{https://planobs.jive.eu/}}.

\subsection{Radio Telescope}
\label{sec:radio_telescope}
\begin{figure} [ht]
   \begin{center}
   \begin{tabular}{cc} %% tabular useful for creating an array of images 
   \includegraphics[height=6cm]{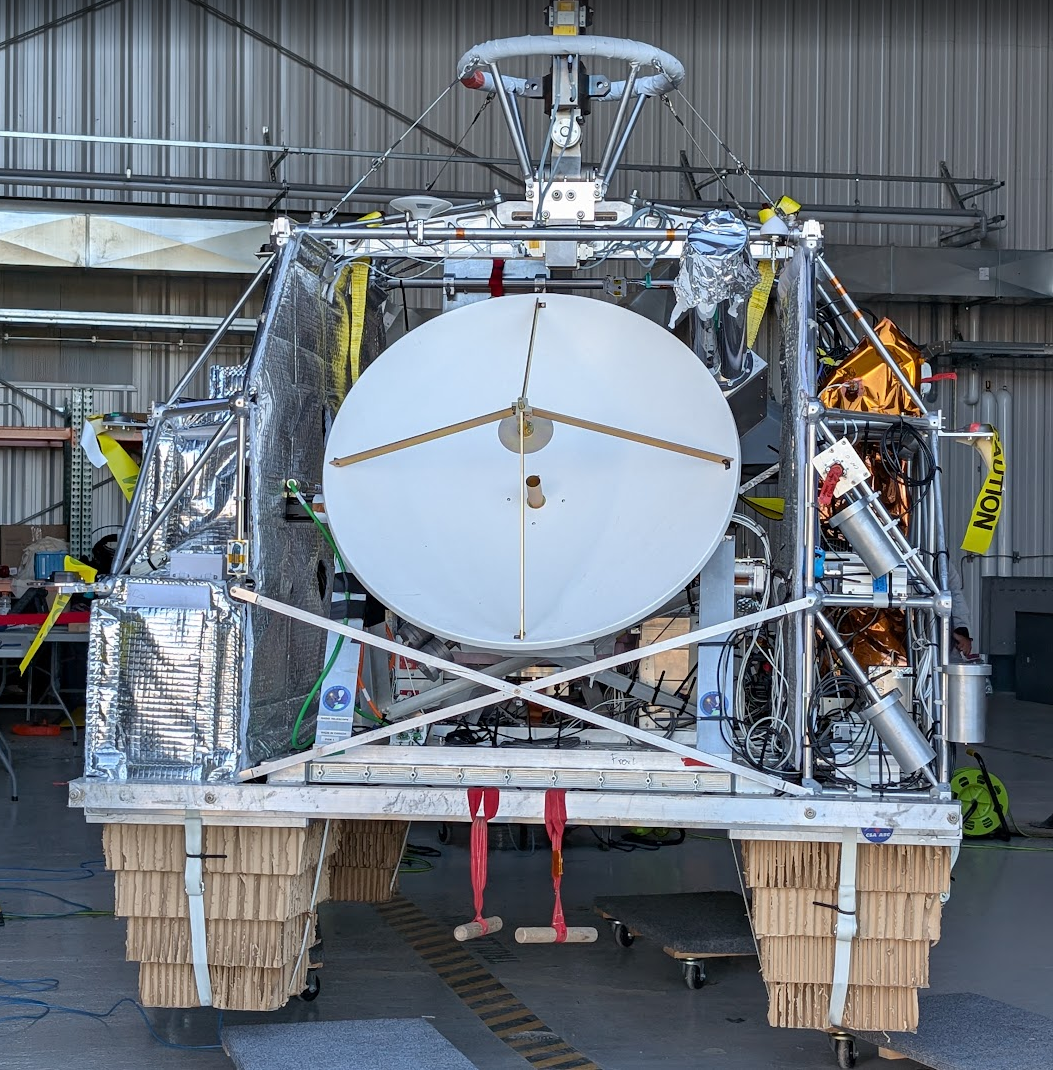} &
   \includegraphics[height=5cm]{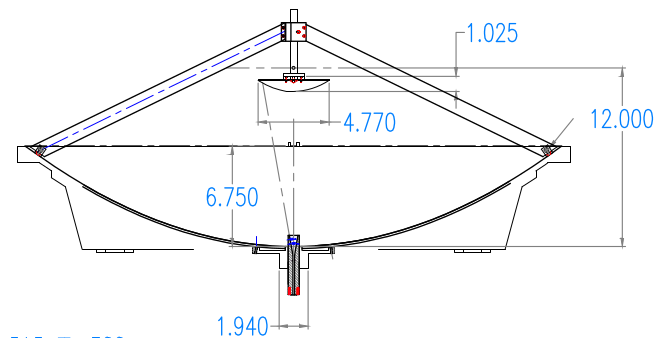}
   \end{tabular}
   \end{center}
   \caption[example] 
%>>>> use \label inside caption to get Fig. number with \ref{} 
   { \label{fig:BVEX_tel} \textit{Left:} The BVEX telescope fully integrated on the CSA/CNES gondola during pointing tests in the highbay at the balloon base in Timmins Ontario. The gondola deck the telescope was mounted to is about 1.1 m wide restricting the width of the mount to about 90 cm. \textit{Right}: Dimensional drawing of the fiberglass dish provided by Mi-Wave, all dimensions are in inches. The dish is 36 inches in diameter.}
   \end{figure} 

As shown in Fig.~\ref{fig:BVEX_tel} the BVEX experiment is part of a shared gondola provided by the CSA and CNES and the size of the telescope is constrained by the inner gondola dimensions of 1.2 m (H)x 1.1 m (W) x 2 m (L). This naturally limited the diameter of the radio dish to 36 inches (91.44 cm) to allow for additional room to route cables as well as for the elevation motor. We therefore used an Mi-Wave 223E-36 radio dish made from fiberglass with integrated stainless steel reinforcement for mounting. The dish itself uses a Cassegrain setup with a primary focal ratio of 0.3, an effective collecting area of about 0.43 m$^2$ at 22 GHz as well as a beamsize of about 1 degree (see rightmost panel in Fig.~\ref{fig:BVEX_tel} for detailed dimensions). The dish is then mounted to the elevation gimbal of the telescope mount using 4 bolted fixtures at the back.

Because the gondola was provided by CSA/CNES the telescope did not need an azimuth pointing system, and only needed to point in elevation. The design and geometry for the telescope mount itself was inspired by both BLAST-TNG\cite{2018SPIE10700E..22L} and EBEX\cite{2018ApJS..239....9E}. The key difference to these designs is that the BVEX telescope mount is entirely free of welds, allowing for disassembly and modification for future balloon missions. Furthermore the telescope was designed in such a way that most of the parts can be fabricated from commercially available Aluminum 6061-T6 stock without the need for material with custom dimensions. 

The outer frame of the telescope mount (i.e. the part that mounts to the gondola) made use of an A-frame geometry which was chosen due to its high strength-to-weight ratio and ease of fabrication. Two diagonal trusses were included at the back of the outer frame to prevent any sort of lateral motion of each A-frame structure. The entire mount was bolted to the gondola using the 10 cm x 10 cm diagonal bolt pattern on the gondola deck. This bolt pattern restricted the width of the telescope mount to 90 cm.
\begin{figure}[ht]
    \begin{center}
   \begin{tabular}{c} %% tabular useful for creating an array of images 
   \includegraphics[height=10cm]{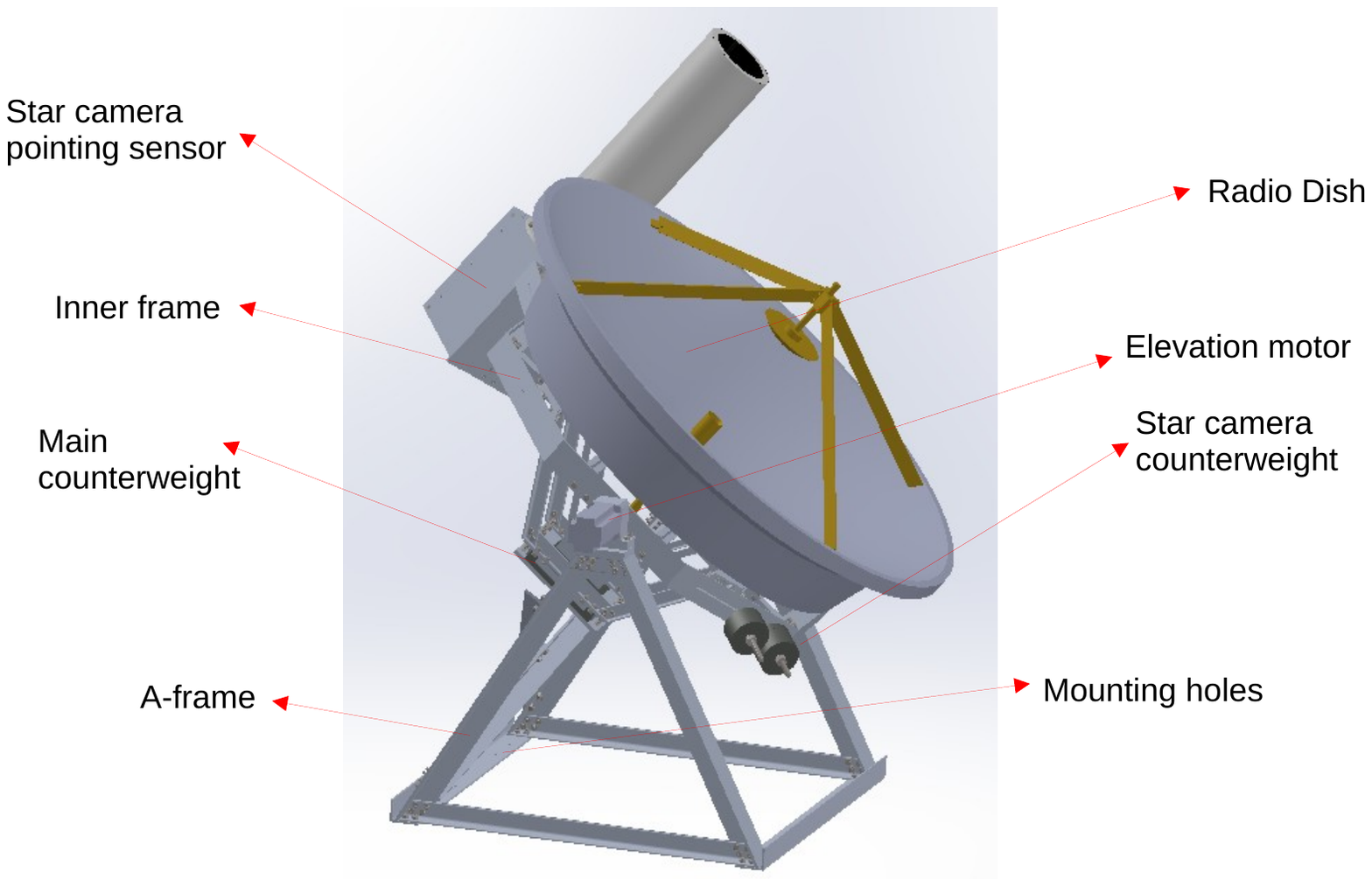} 
   \end{tabular}
   \end{center}
   \caption[example] 
%>>>> use \label inside caption to get Fig. number with \ref{} 
   { \label{fig:BVEX_CAD} Annotated CAD model of the BVEX telescope. The main A-frame structure serves as a base for the rotating octagonal inner frame/elevation gimbal. All joints are weld-free and the main structural joints are over-constrained using 4 M6 bolts. Two separate counterweights balance out the star camera and the radio dish. The entire telescope weighs about 50 kg and is bolted to the gondola using the mounting holes at the bottom of the telescope.}
\end{figure}

For the inner elevation gimbal an octagonal geometry was chosen to keep the telescope weight at a minimum. Unlike BLAST\cite{2018SPIE10700E..22L} the BVEX telescope does not have a cryostat, we therefore included a counterweight at the back of the inner elevation gimbal to balance out the weight of the telescope dish. Due to the fact that the telescope mount is narrower than the dish diameter, the dish is mounted in front of the elevation axis. Because of the maximum height constraint of 1.2 m on the gondola the star camera pointing sensor had to be mounted on one of the diagonal sides of the inner elevation gimbal, which therefore necessitated a second counterweight on the side opposite to the star camera. A CAD model of the entire telescope can be seen in Fig.~\ref{fig:BVEX_CAD}.

 \begin{table}[ht]
\caption{Minimum CNES safety margins as determined through finite-element-analysis of the telescope mount and dish when subject to the maximum expected accelerations present during separation (7.2g vertical, 1.4 g horizontal) for different impact angles in the horizontal plane. Margins are positive if the safety requirement is met. For each load case and component we quote two margins one with respect to the ultimate strength and the other with respect to the yield strength of the materials involved. Note that the minimum margins tend to be underestimated by the analysis software and occur around the bolted fixtures.} 
\label{tab:margins}
\begin{center}       
\begin{tabular}{|l|l|l|l|l|l|l|}
\hline
\rule[-1ex]{0pt}{3.5ex}  &\multicolumn{2}{|c|}{Outer Frame} &  \multicolumn{2}{c|}{Gimbal} &  \multicolumn{2}{c|}{Dish} \\
\hline
\rule[-1ex]{0pt}{3.5ex}   Angle &Yield & Ultimate & Yield & Ultimate & Yield & Ultimate   \\
\hline
\rule[-1ex]{0pt}{3.5ex}  0 &1.80&1.6&0.22&0.15&4.18&3.87\\
\hline
\rule[-1ex]{0pt}{3.5ex}  45 &1.24&1.1&0.31&0.24&4.18&3.86\\
\hline
\rule[-1ex]{0pt}{3.5ex}  90 &1.11&0.99&0.66&0.57&4.11&3.8\\
\hline 
\end{tabular}
\end{center}
\end{table}

Apart from the form factor the other design constraint was given by the accelerations that the payload must endure during separation, that is, the end of the balloon flight when the gondola is separated from the balloon and the parachute opens. As per the CNES requirements the entire telescope had to be designed for 7.2g of vertical acceleration and 1.4g of horizontal acceleration. The design was tested under three load cases (assuming the telescope is pointed at the horizon in its storage position) where the horizontal component of the acceleration vector is directed perpendicular to the radio dish (0 deg) , parallel to the radio dish (90 deg) or diagonally at the dish (45 deg) using finite-element-analysis (FEA). This analysis was performed separately on the outer frame, the elevation gimbal as well as the dish itself. CNES safety margins comparing against yield and ultimate strengths can be found in Tbl.~\ref{tab:margins}.

\subsection{Receiver}
\label{sec:receiver}
\begin{figure}[ht]
    \begin{center}
    
   \begin{tabular}{cc} %% tabular useful for creating an array of images 
   \includegraphics[height=9cm]{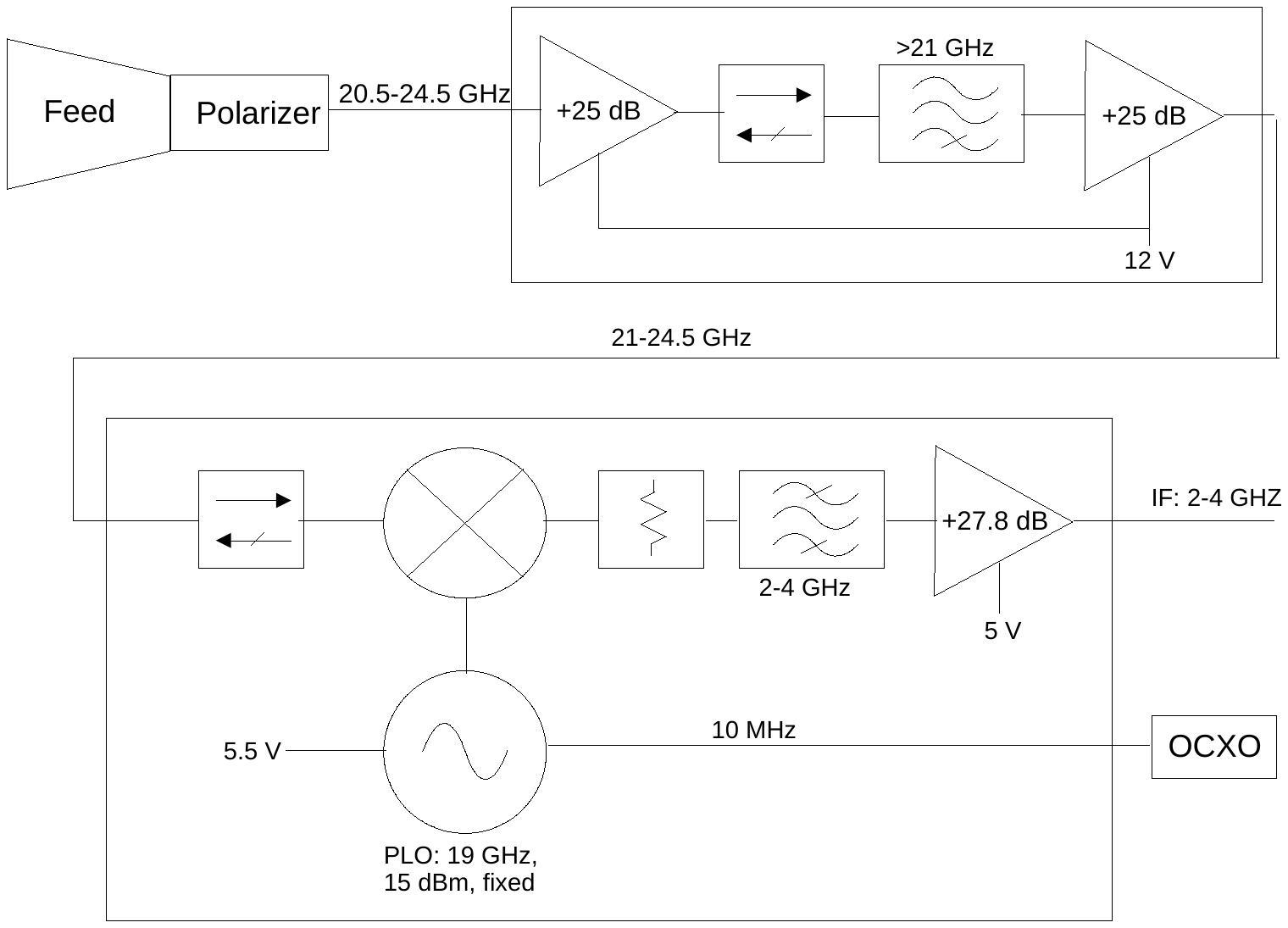} &
   \includegraphics[height=9cm]{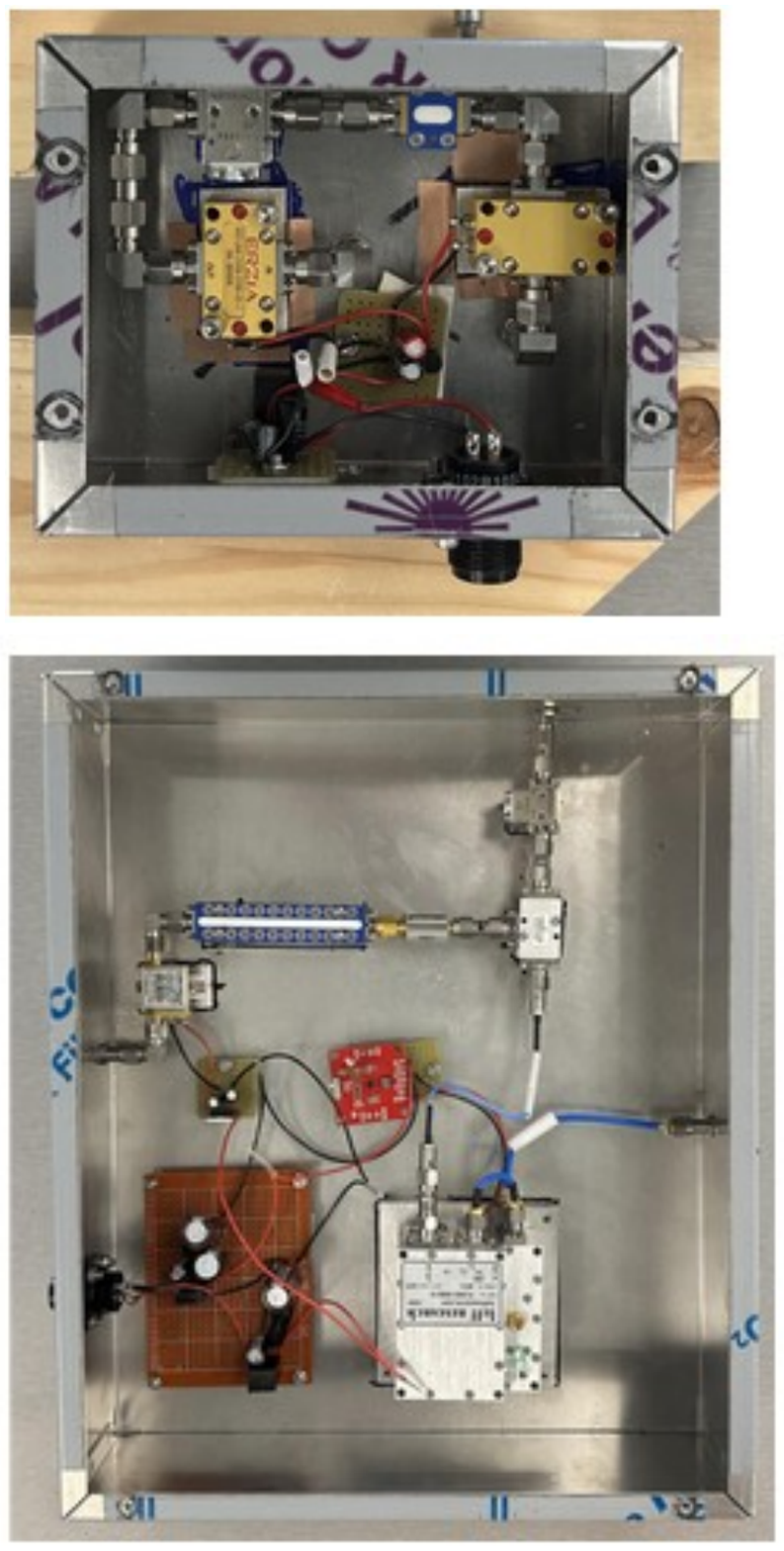}\\
   \end{tabular}
   \end{center}
   \caption[example] 
%>>>> use \label inside caption to get Fig. number with \ref{} 
   { \label{fig:BVEX_RX} \textit{Left:} Block diagram of the ambient temperature BVEX receiver from Ref. \citenum{BVEXInstrument}. The 20.5-24.5 GHz signals from the polarizer are amplified and signals $<21$ GHz are rejected at the high-pass filter in between the two LNAs. The remaining signals are then mixed with a 19 GHz phase-locked (PLO) local oscillator and are filtered to 2-4 GHz making the entire system sensitive to 21-23 GHz signals. Not all losses are shown in the diagram. \textit{Top Right:} Low noise amplifiers in their enclosure which mounted to the back of the telescope. \textit{Bottom Right:} Fully integrated mixing chain inside its enclosure which is mounted inside a pressure vessel.}
\end{figure}
In order to minimize the data volume and cost, the BVEX telescope used a single polarization and single sideband heterodyne receiver built from commercial off-the-shelf components as can be seen in Fig.~\ref{fig:BVEX_RX}. To amplify the signal coming from the antenna we used two room-temperature ERZ-LNA-2100-2700-25-2 Low Noise Amplifiers (LNA) each with a noise figure of 2 dB and a gain of 25 dB. In between the two amplifiers we placed a 21 GHz high-pass filter to reject the unwanted sideband. The 21-24.5 GHz signals are then fed into a mixer where they are mixed with a 19 GHz local oscillator (Luff Research PLDRO-19000-10) and subsequently filtered by a 2-4 GHz bandpass filter. All in all this makes the receiver sensitive to a 22 GHz central frequency with a 2 GHz bandwidth. The 2-4 GHz intermediate frequency(IF) signals are then sampled in the second Nyquist zone by the RFSoC backend which is discussed in Ref. \citenum{Bagchi2026bvex}. This scheme was adopted to improve rejection of the unwanted sideband at 15-17 GHz. In order to minimize the noise temperature the LNAs were kept right behind the antenna feed of the telescope. The mixing chain needed to operate at atmospheric pressure and was therefore located inside a pressure vessel. More information on the receiver can be found in Ref. \citenum{BVEXInstrument}.

The receiver had a total gain of about 60 dB as well as a design noise temperature of 367 K (depending on the temperature of the first stage of the receiver in the stratosphere). The SEFD of BVEX alone therefore was $2.4 \cdot 10^6$ Jy, baseline sensitivities in Jy$\cdot$s$^{1/2}$ when combined with existing ground-based facilities can be found in Tbl.~\ref{tab:sensitivity}

\begin{table}[ht]
\caption{Baseline sensitivities when combining BVEX with existing ground-based observatories. We assume a bandwidth of 512 MHz for all observatories except the Haystack 37m where we assume a 2 GHz bandwidth. Furthermore we adopt $\eta_q = 0.88$ to account for requantization to 2-bits } 
\label{tab:sensitivity}
\begin{center}       
\begin{tabular}{|l|l|l|l|l|}
\hline
\rule[-1ex]{0pt}{3.5ex}   Partner Station(s)&GBT &Effelsberg & VLBA & Haystack\\
\hline
\rule[-1ex]{0pt}{3.5ex}  $\sigma_{rms}$&0.2 Jy $\cdot$s$^{1/2}$ & 0.5 Jy $\cdot$s$^{1/2}$&1.4 Jy $\cdot$s$^{1/2}$&0.9 Jy $\cdot$s$^{1/2}$\\

\hline 
\end{tabular}
\end{center}
\end{table}

\subsection{Pointing}
\label{sec:pointing}
\begin{figure}[ht]
    \begin{center}
   \begin{tabular}{c} %% tabular useful for creating an array of images 
   \includegraphics[height=10cm]{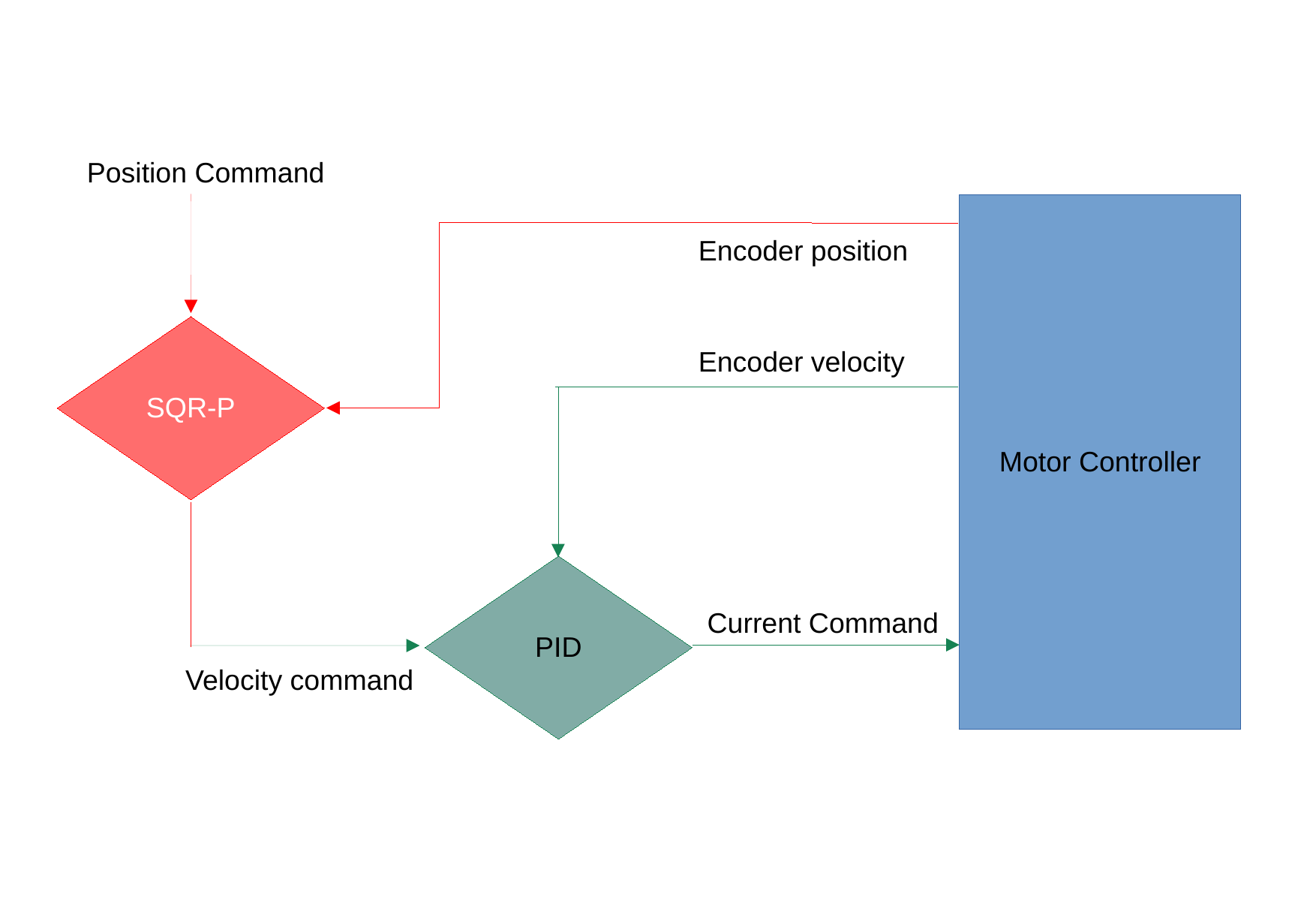} 
   \end{tabular}
   \end{center}
   \caption[example] 
%>>>> use \label inside caption to get Fig. number with \ref{} 
   { \label{fig:control} Block diagram of the BVEX pointing control loop. For the velocity loop only the green parts of the diagram would be active, which involved a simple PID loop to command motor current. For the position loop both the red and green parts would be active and the square-root proportional (SQR-P) loop would command the velocity PID.}
\end{figure}

As mentioned in Sec.~\ref{sec:radio_telescope} the pointing in azimuth was provided by CSA/CNES by rotating the balloon gondola using the pivot which connects the gondola to the rest of the flight train and balloon. The telescope mount therefore only had to point in elevation. For elevation control we used a Kollmorgen  AKM24F-ACBNAA-0 brushless DC-motor (BLDC) with a built in single-turn absolute BiSS sine encoder for feedback. The BLDC option was chosen to mitigate vibrations in the telescope mount during operation and similar models have flight heritage through missions like BLAST\cite{2018SPIE10700E..22L}. To get an absolute estimate of the pointing we used the \texttt{blastcam} star camera design\footnote{\url{https://github.com/BlastTNG/blastcam}} we however ran the camera software as part of the BVEX Control Program (\texttt{bcp})\footnote{\url{https://github.com/fissellab/bcp}} on the main flight computer to reduce power consumption and star camera weight. Once integrated on the gondola the telescope had a pointing range in elevation of 18-55 degrees. This range was limited on the lower end by a structural beam blocking the telescope aperture and as a result distorting the beam shape of the telescope and on the upper end by another structural component interfering with the star camera baffle. Both telescope pointing as well as star camera images and solutions were sent to the ground-station through the CSA/CNES S-band telemetry downlink. 

The motor control software in \texttt{bcp} was inspired by the Master Control Program (\texttt{mcp})\footnote{\url{https://github.com/BlastTNG/flight}} used for BLAST and consisted of two control modes: position and velocity control shown in the block diagram in Fig.~\ref{fig:control}. For the velocity mode we employed a simple PID loop that would command the current (and hence the torque) of the motor and use the encoder velocity as a feedback to keep the velocity at the desired setpoint. For the position mode we employed a second control mode that used a loop proportional to the square-root of the position error to command velocity, this velocity would in turn be fed into the velocity control PID loop to command the motor torque.

\begin{figure}[ht]
    \begin{center}
   \begin{tabular}{c} %% tabular useful for creating an array of images 
   \includegraphics[height=8cm]{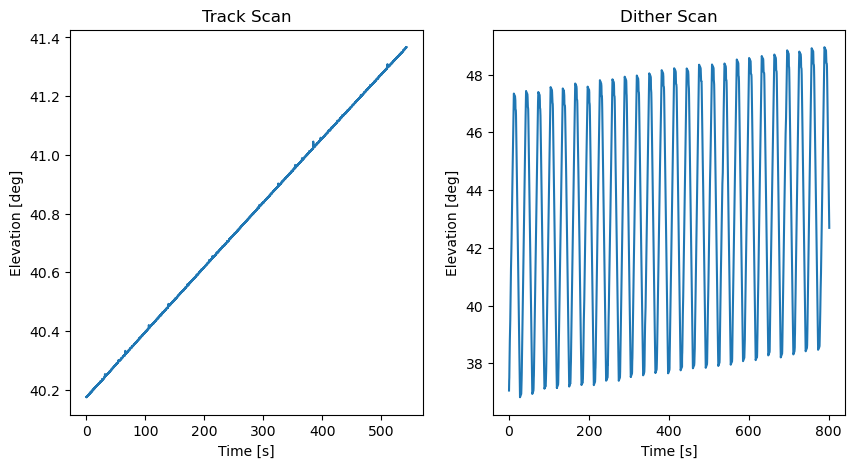} 
   \end{tabular}
   \end{center}
   \caption[example] 
%>>>> use \label inside caption to get Fig. number with \ref{} 
   { \label{fig:scan} Elevation encoder readings from the two most common scan modes. The track scan (\textit{Left}) follows the elevation of a target with fixed right ascension (RA) and declination (DEC) the dither scan (\textit{Right}) slews the telescope across the target's elevation while the scan center tracks the rising or setting of the target.}
\end{figure}

BVEX had two main scan modes: Tracking mode and dither mode. The tracking mode requires following the elevation of a target at fixed right ascension and declination. This mode was used for both spectral line observations as well as VLBI observations. The dither mode was implemented for ground based testing and involved scanning across the source while the scan center tracked the source in elevation. By keeping the azimuth fixed and allowing the source to drift past the azimuth the telescope was pointed at, it was possible to make elevation drift scans of a target. This dither scan mode was used for beam-map measurements using either the Sun or the Moon as a target (both subtend a smaller angular size than the FWHM of the BVEX telescope beam). The track scan had a pointing accuracy of 0.01 deg which is well within the requirement for BVEX of 1/10th of the beam FWHM (0.1 deg). Examples of both of these scans can be seen in Fig.~\ref{fig:scan}.

\section{POSITION TRACKING}
\label{sec:position}
One challenge that will arise when we try to achieve VLBI from a balloon is the fact that the balloon telescope moves in an unconstrained and unpredictable way. The coherence time of the observation therefore is not only set by the jitter of the on-board clock, but is also set by the ability to reconstruct the position. This measurement is needed to apply corrections to the geometric time delay between the balloon and the ground station despite the change in the balloon position by many wavelengths (about 1000 wavelengths in 1s at 22 GHz). In order to keep the phase error along the telescope line-of-sight to less than 1 radian we need to know the line-of-sight component of the position vector to a precision of
\begin{equation}
    \delta|\vec{r}'| = \frac{\lambda}{2\pi} = \frac{c}{2\pi \nu}
\end{equation}
where $\lambda$ is the observing wavelength, $c$ is the speed of light and $\nu$ is the observing frequency. For the design of the position tracking system we adopted a margin of about 40\% and therefore used the requirement $\lambda/10$\cite{BVEXInstrument}. At K-Band (22 GHz) we therefore have a position tracking requirement of at least 1.3 mm.

\begin{figure}[ht]
    \begin{center}
   \begin{tabular}{c} %% tabular useful for creating an array of images
   \includegraphics[height=5cm]{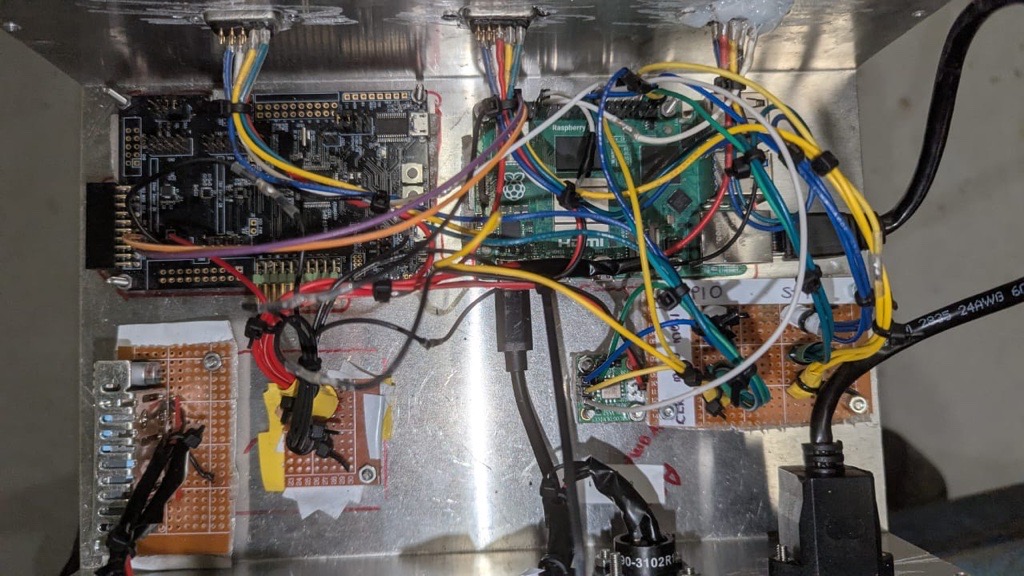}\\
   \includegraphics[height=7cm]{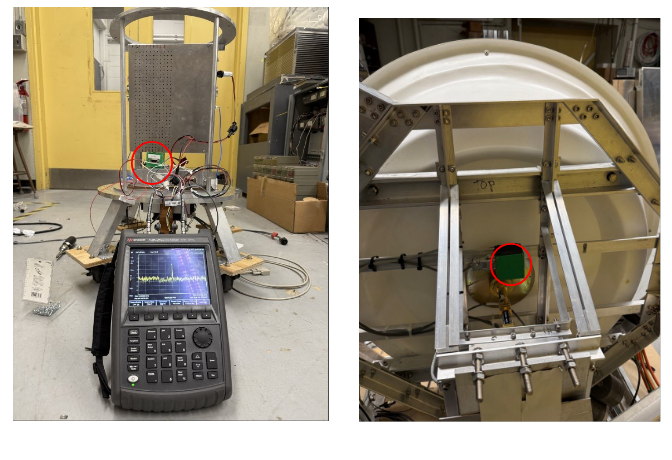}
   \end{tabular}
   \end{center}
   \caption[example] 
%>>>> use \label inside caption to get Fig. number with \ref{} 
   { \label{fig:pos-box} \textit{Top:} Position tracking box containing one of the accelerometers, a 3-axis gyro as well as a Raspberry Pi 5 for readout. The box was mounted on one of the side compartments of the CSA gondola \textit{Bottom:} Location of the other two accelerometers circled in red one was placed inside the pressure vessel on top of the crystal oscillator enclosure (\textit{Left}), the other one was placed behind the antenna feed of the telescope (\textit{Right}).}
\end{figure}

Conventional GPS units only have an accuracy of about 0.1 $-$ 1 m and sample at a cadence that is too low to capture position changes on short timescales of $<$1s. BVEX therefore is using a VEGA 40 GPS board to return the absolute position and velocity on long timescales $>$1s and 3 Analog Devices ADXL-355Z 3-axis accelerometers to characterize the jitter on short $\ll$1s timescales. For a perfectly temporally decorrelated accelerometer with infinite resolution, the velocity error in a single axis as a function of integration time is given by
\begin{equation}\label{eq:deltav}
    \delta v = \sigma_{rms}\sqrt{\frac{\tau}{f_{smp}}}
\end{equation}
where $\sigma_{rms}$ is the noise rms of the accelerometer, $t$ is the integration time and $f_{smp}$ is the sampling rate. Similarly, the position error is given by
\begin{equation}\label{eq:deltax}
    \delta x = \frac{\sigma_{rms}}{\sqrt{3}}\sqrt{3\left(\frac{\tau}{f_{smp}}\right)^2+\frac{\tau^3}{f_{smp}}-\frac{\tau}{f_{smp}^3}}.
\end{equation}
A detailed derivation of Eqs.~\ref{eq:deltav} and ~\ref{eq:deltax} can be found in App.~\ref{app:derivation}. Note that the sampling rate here plays two crucial roles, on one hand it reduces the cumulative error in position and velocity, and on the other it ensures that the acceleration changes are Nyquist sampled. For BVEX the accelerometers were sampling at 1000 Hz with a noise floor of about 2.4-3.5 mm/s$^2$.

Any long time-scale drifts induced by the inaccuracy of the GPS positon can be fitted for during correlation as those errors are expected to be mostly constant over the integration time. To ensure redundancy the three accelerometers were mounted at different locations on the gondola: one right behind the antenna feed, another inside the pressure vessel housing the crystal oscillator clock and the mixing chain, and one inside a sheet-metal enclosure which included the position sensor readout (a Raspberry Pi 5) as well as a 3-axis gyroscope (DK-2038HT). The locations of the different sensors are labeled in Fig.~\ref{fig:pos-box}. 

\begin{figure}[ht]
    \begin{center}
   \begin{tabular}{cc} %% tabular useful for creating an array of images
   \includegraphics[height=6cm]{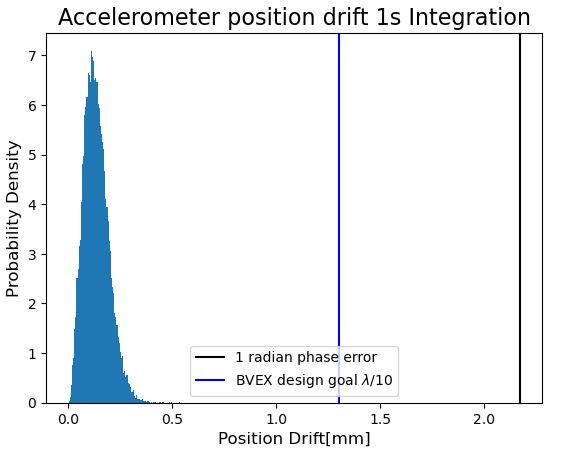}&
   \includegraphics[height=6cm]{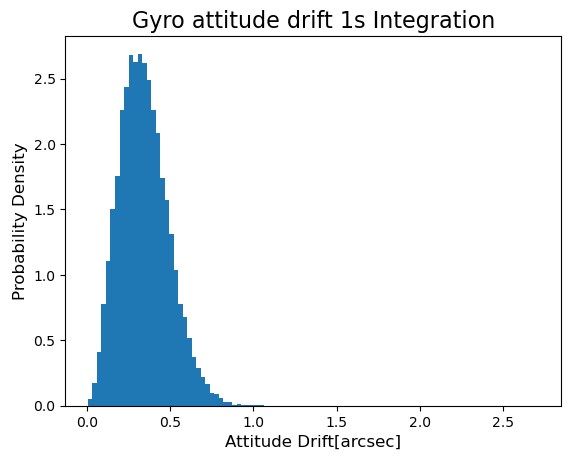}
   \end{tabular}
   \end{center}
   \caption[example] 
%>>>> use \label inside caption to get Fig. number with \ref{} 
   { \label{fig:pos-drift} \textit{Left:} Histogram of  accelerometer position drifts ($\delta|\vec{r}|$) over 1s time scales. The lines indicate the position-tracking requirement for BVEX as well as the requirement for a 1 radian phase error. \textit{Right:} Attitude drifts of the gyroscope over 1 s time scales ($\delta\alpha$). Each 1s time chunk was mean-subtracted individually, to remove long timescale drifts of the gyroscope.}
\end{figure}

The gyroscope measures the angular velocity of the telescope and was included to account for changes in the gondola attitude at high cadence. The attitude is needed to correct for the fact that the accelerations are measured in the gondola's reference frame and not in the Earth-Centered-Earth-Fixed (ECEF) frame which is the coordinate system of choice for VLBI. The precision requirement for the attitude arises from the dot product between the telescope line-of-sight displacement vector and the accelerometer displacement vector. 
\begin{equation}
    |\vec{r}'| = |\vec{r}|\cos(\alpha)
\end{equation}
where $|\vec{r}'|$ is projection of the accelerometer displacement vector $|\vec{r}|$ onto the line-of-sight and $\alpha$ is the angle between the two vectors. We can use this to derive the position error along the line of sight in terms of the accelerometer position error and the attitude error
\begin{equation}
    \delta|\vec{r}'| = \sqrt{\left(|\vec{r}|\sin(\alpha)\delta\alpha\right)^2+\left(\cos(\alpha)\delta|\vec{r}|\right)^2}.
\end{equation}
We can see that there are two limits: If $\alpha = 0^{\circ}$ the accelerometer is moving parallel to the line-of-sight where $\delta|\vec{r}'| =\delta|\vec{r}|$ as expected. If $\alpha = 90^{\circ}$ we get
\begin{equation}
   \delta\alpha = \frac{\delta|\vec{r}'|}{|\vec{r}|} = \frac{\lambda}{10|\vec{r}|}
\end{equation}
where the last step follows from the position tracking requirement that $\delta|r| \leq \lambda/10$. The quantity $|\vec{r}|$ is the displacement of the accelerometer over the time scale that we measure the position drift. Stratospheric balloons typically have velocities of about 10 m/s and over a 1 s integration time $|\vec{r}|=10$\,m is typical. Therefore, for a wavelength of 1.3 cm we need to determine the attitude to a precision of 26 arcseconds. The gyroscope readings can then be combined with data from other auxiliary sensors such as the star camera, an optical wide-field camera that takes pictures of the sky and compares the observed stars to a reference catalog in order to solve for the camera attitude, the onboard CNES Inertial measurement Unit (IMU), the GPS unit as well as the CSA's $\mu$PRISM sensor package to give a complete attitude reconstruction.

To get estimates for $\delta|\vec{r}|$ as well as $\delta\alpha$ we conducted an 8 hour stationary test where we collected data for one of the accelerometers as well as the gyroscope. We then split the time series data into 1\,s chunks (the coherence time assumed for sensitivity calculations). For the accelerometer data we subtracted a global mean and applied a double trapezoid integration to get the position drift. Similarly, for the gyroscope we did a single integration on each 1\,s chunk however we subtracted the mean on a chunk-per-chunk basis to remove drifts in the gyroscope on $>1$ s time scales. Histograms of all 1\,s chunks can be seen in Fig.~\ref{fig:pos-drift}. We therefore report a mean position drift of about 0.2 mm on 1\,s time scales and mean attitude drift of 0.4 arcsec on 1\,s time scales. 

An aspect that remains to be investigated here is how these drifts can be applied to real interferometric data and how they can be ingested by a software like DiFX \cite{2007PASP..119..318D} which is the standard correlator software for VLBI or a modified version thereof. One could even imagine specifying the position estimate from the position tracking sensors as an initial guess to find interference fringes on the first couple of time chunks of VLBI data. Once fringes have been found for these time chunks we can then use the position estimate from that correlation as an initial guess for the next time chunk. Future ground-based VLBI tests of a next-generation BVEX mission will give more insights into this matter.

\section{GROUND-BASED TESTING, THE 2025 TIMMINS CAMPAIGN, AND FUTURE PROSPECTS}
\label{sec:ground_tests}
Due to the high noise temperature and small diameter of the radio dish only the brightest radio sources can be used to test instrument performance in single dish mode. Initial tests involved either mapping of bright continuum sources like the Sun and the Moon to characterize the antenna beam shape, or observations of bright $10^4-10^6$ Jy 22.235 GHz water masers to test the on-board spectrometer. In this section we will primarily focus on maps of the moon made in Timmins, Ontario on August 18th 2025 starting at 09:30 UTC, the spectroscopic tests are covered in Ref.~\citenum{Bagchi2026bvex}. 

\begin{figure}[ht]
    \begin{center}
   \begin{tabular}{c} %% tabular useful for creating an array of images
   \includegraphics[width = \linewidth]{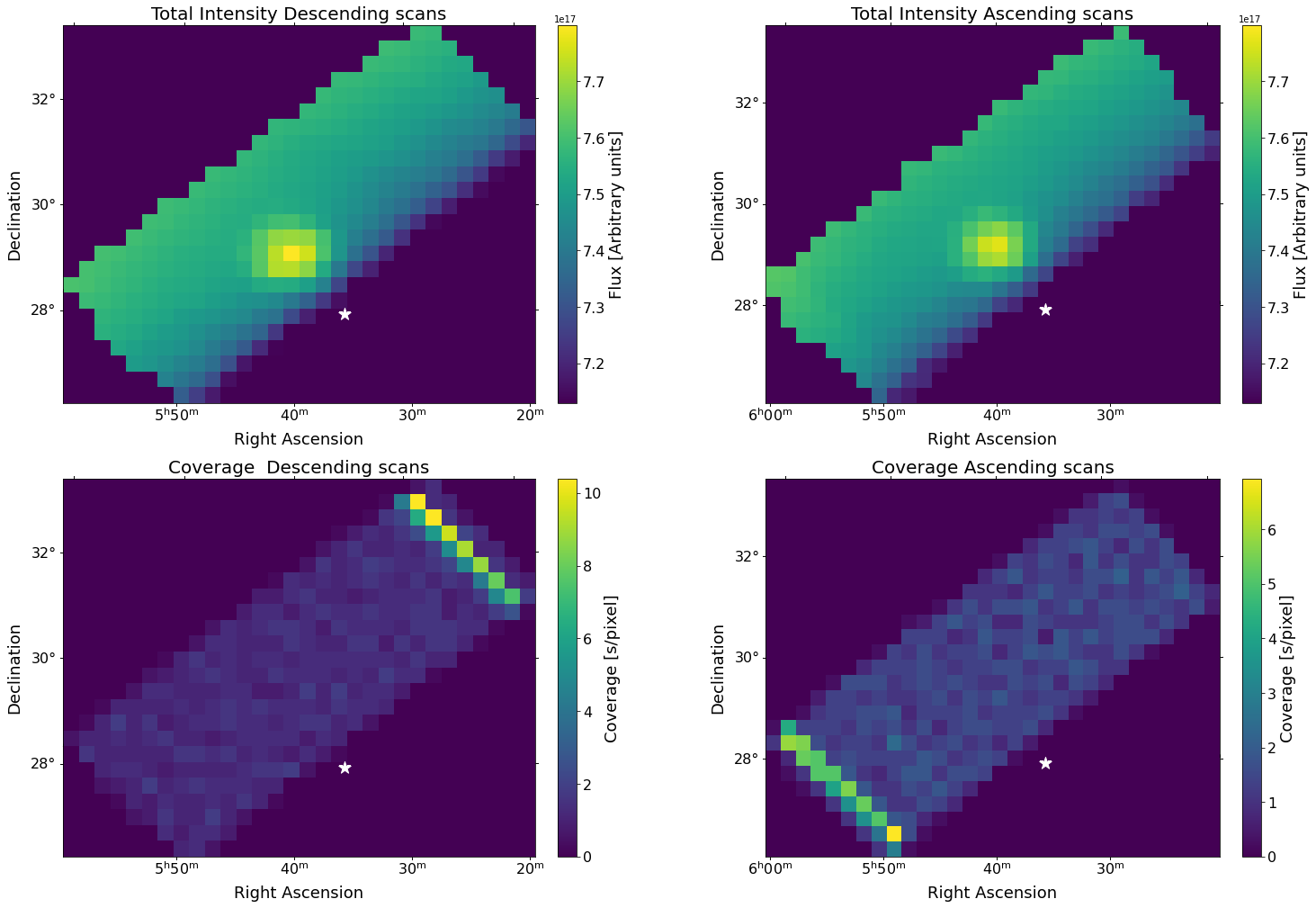}
   \end{tabular}
   \end{center}
   \caption[example] 
%>>>> use \label inside caption to get Fig. number with \ref{} 
   { \label{fig:moonmap} Ground-based BVEX map of the Moon. The ascending (right column) and descending scans (left column) were reduced separately to test for backlash in the motor coupling. The bottom row shows the map coverage in each case in seconds/pixel. The white star in the map marks the true location of the Moon. Note that the elevation direction is along the long edge of the map and the azimuth along the short edge. Due to inaccuracies of the GPS heading the pointing offset in the azimuth direction is significantly larger than in the elevation direction.}
\end{figure}

The Moon was chosen as a target because it is bright (a brightness temperature of 239 K, during full moon phase\cite{2021IGRSL..18.2021Y}) and still is a point source (angular size of about 30 arcmin) compared to the 1$^{\circ}$ beam of the BVEX telescope. It can therefore be used to characterize the beam-shape, pointing offset, and any sort of backlash in the elevation motor control system. Using the dither scan from Sec.~\ref{sec:pointing} we made ground-based elevation drift maps of the moon as seen in Fig.~\ref{fig:moonmap} using a naive map-maker (i.e. binning raw fluxes into pixels). Note that these observations were made at sunrise and the gradient in the map is due to the atmosphere warming up during the observations. We separated ascending and descending scans to test for backlash in the elevation motor coupling. Because the observations were made at dawn we could not use the star camera to measure the attitude of the telescope. Instead we used the elevation encoder reading to estimate elevation, and the GPS heading to measure the azimuth. While the encoder is extremely precise (0.01$^\circ$ precision) the GPS readings have an uncertainty of 0.08 deg provided that the GPS receiver has a good line of sight to multiple GNSS satellites, which was not a given for these observations due to the proximity to the highbay at the balloon-base. As a result we measured a significant pointing offset between the telescope and the GPS heading in the azimuthal direction. During flight however, the azimuth pointing would be controlled by the CNES/CSA who use an on-board and more accurate Inertial Measurement Unit (IMU) unit with 1 arcmin precision as feedback. 

\begin{figure}[ht]
    \begin{center}
   \begin{tabular}{cc} %% tabular useful for creating an array of images
   \includegraphics[height=5.5cm]{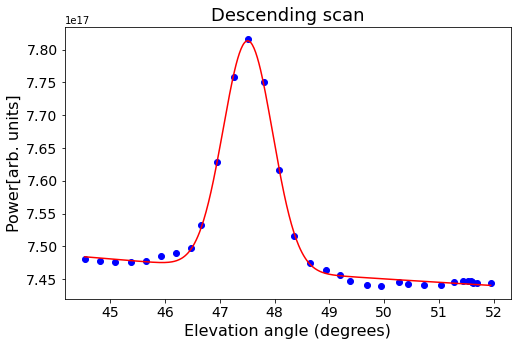}&
   \includegraphics[height=5.5cm]{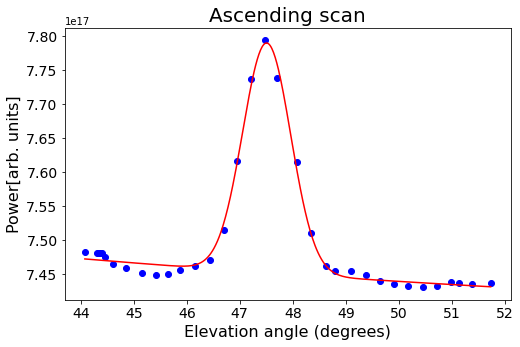}
   \end{tabular}
   \end{center}
   \caption[example] 
%>>>> use \label inside caption to get Fig. number with \ref{} 
   { \label{fig:beam-fit} \textit{Left:} Beam fit of the descending scan giving an elevation pointing offset of 0.21 deg in elevation \textit{Right:} Beam fit of the ascending scan giving a pointing offset of 0.18 deg in elevation. Both datasets gave a beam FWHM of 1.06 deg.}
\end{figure}

To determine the size of the antenna beam from this data we chose the scan with the brightest peak value and took this to be the scan where the Moon was at the same azimuth as the telescope. We then proceeded by fitting a Gaussian super-imposed on a secant curve to this scan to determine the full-width at half-maximum (FWHM) as well as the pointing offset between the encoder and the main telescope beam. The secant curve was included to model the changing atmospheric loading along the scan. Both of these fits can be seen in Fig.~\ref{fig:beam-fit} giving a FWHM of 1.06 degrees in excellent agreement with the optical specifications of the radio dish. In this dataset we obtain very similar pointing offsets for both the ascending (0.21 deg) and descending scans (0.18 deg) and conclude that for this observing mode backlash from the motor coupling was negligible.

\begin{figure}[ht]
    \begin{center}
   \begin{tabular}{cc} %% tabular useful for creating an array of images
   \includegraphics[height=6.5cm]{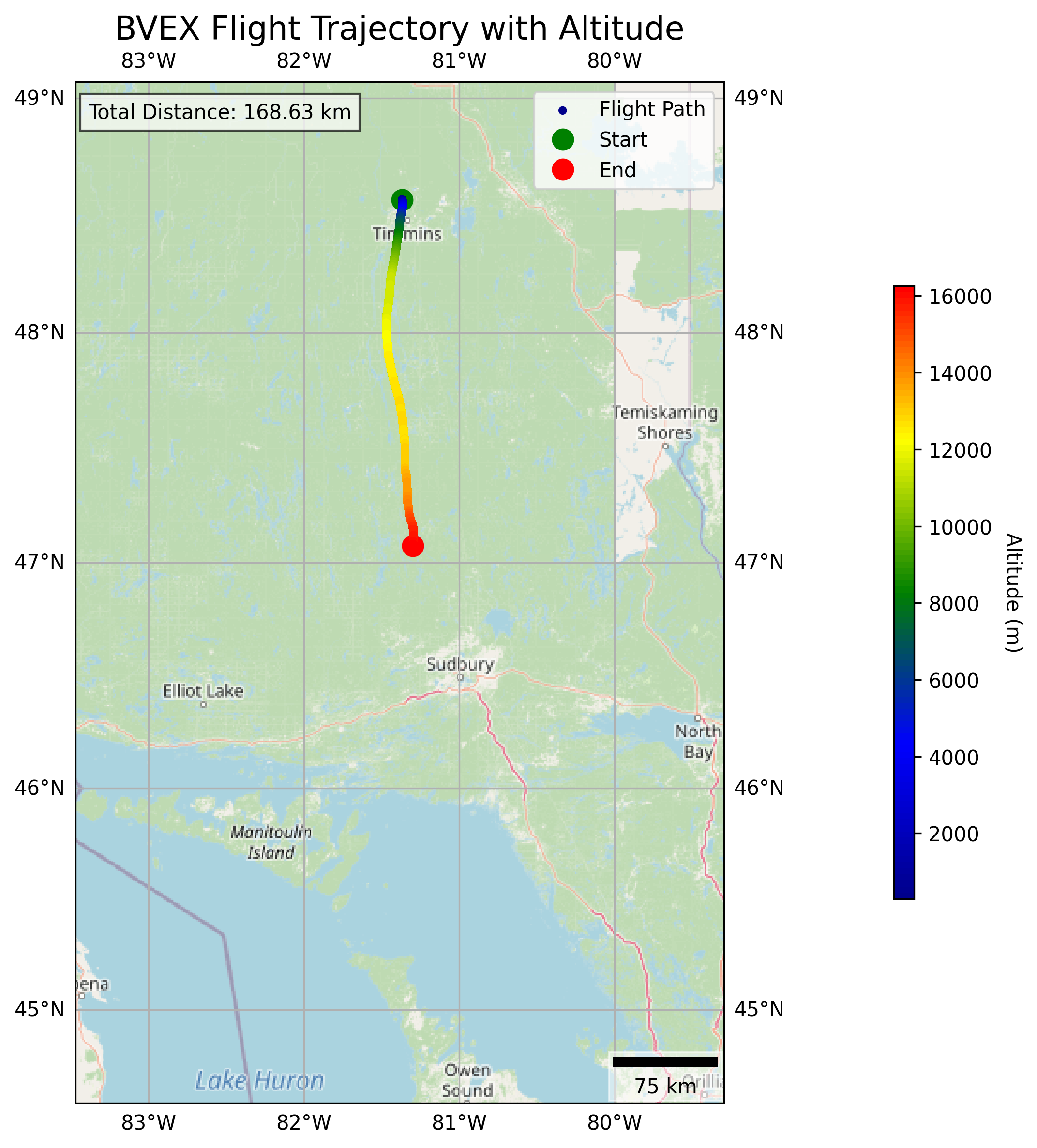}&
   \includegraphics[height=6.5cm]{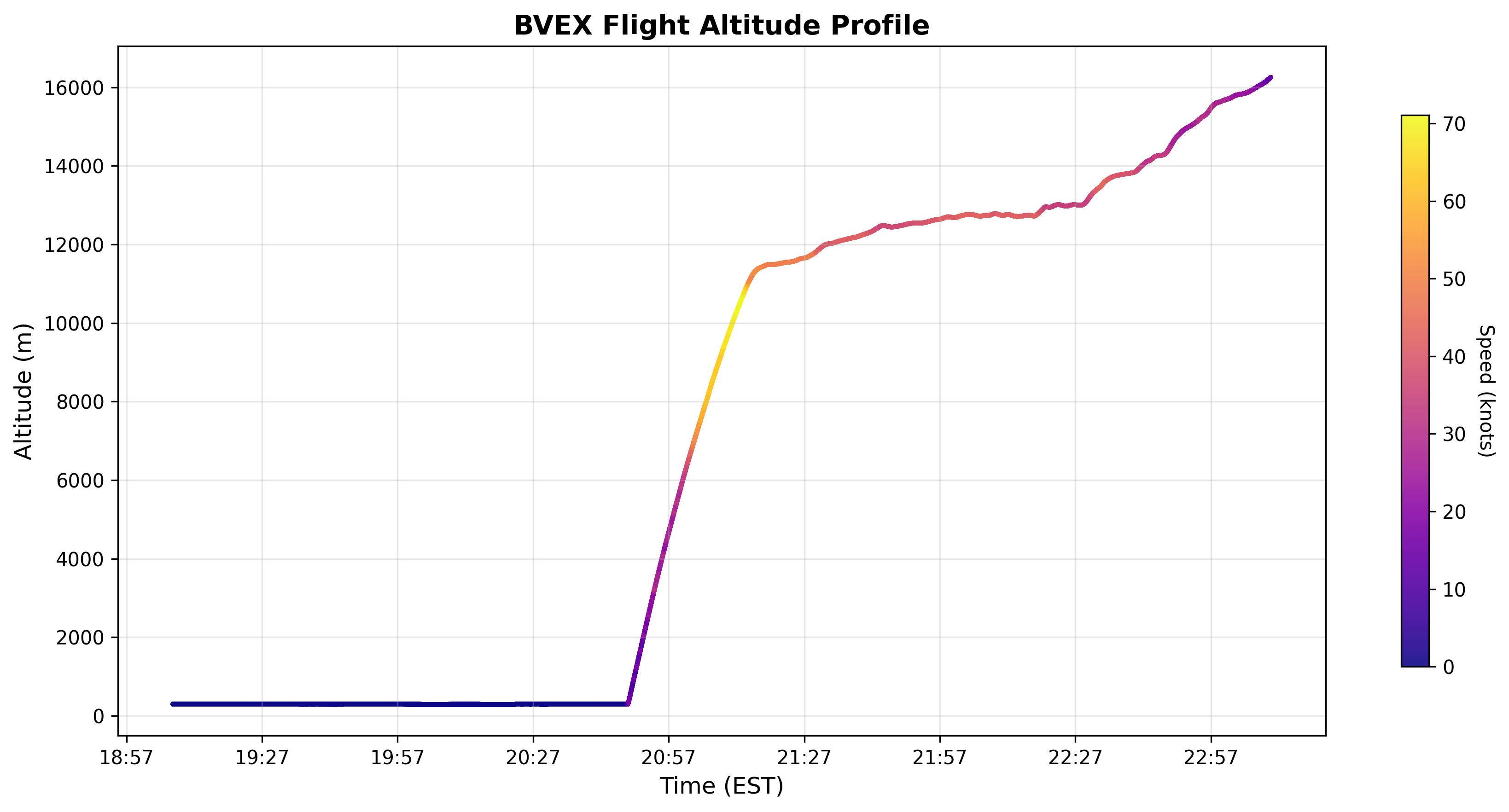}
   \end{tabular}
   \end{center}
   \caption[example] 
%>>>> use \label inside caption to get Fig. number with \ref{} 
   { \label{fig:flight_traj} \textit{Left:} Flight trajectory of the BVEX 2025 flight from Timmins, Ontario. The color scale represents altitude. \textit{Right:} Altitude profile of the BVEX flight with color representing the speed of the payload in knots. Up to 21:15 EST the altitude profile looks nominal before flattening out. At around 22:30 CNES dropped ballast to increase the vertical speed of the payload with minimal effect. }
\end{figure}

The first BVEX flight launched on August 29th 2025 at 20:47 EST from Timmins Airport, Ontario, Canada as part of the 2025 CSA STRATOS campaign. Our team was awarded 4 hours of target of opportunity (TOO) time with the VLBA and the Effelsberg 100m telescope near Bad-Münstereifel, Germany for simultaneous ground-based observations of bright radio galaxies and the massive star forming region W49N. On top of this the Haystack 37-m joined observations as well, to give shorter baseline coverage. Fig.~\ref{fig:flight_traj} shows the trajectory and altitude profile of the BVEX flight. After half an hour the balloon ascent rate dropped substantially. Ballast drops were attempted to get the balloon to ascend to the target altitude (32 km) but after 2 hours it became clear that the balloon had a serious leak and the flight had to be terminated. The flight therefore ended early and the entire gondola landed in a lake near Sudbury, Ontario and we hence did not collect any VLBI data.

Our team has secured additional funding to rebuild BVEX and re-launch the telescope from Palmas, Tocantins, Brazil in late summer 2027. This next-generation BVEX mission will include a number of upgrades especially for the amplification stage of the receiver. This will on one hand involve an improved first LNA as well as a lower-loss polarizer which will improve the receiver noise temperature to 155 K which is less than half the noise temperature presented in this paper. We will also include a noise diode for in-flight noise temperature calibration which has the potential to enable imaging of VLBI data rather than simply finding interference fringes.

\section{CONCLUSIONS}
\label{sec:conclusions}
The Balloon-borne VLBI Experiment (BVEX) is a novel balloon-mission concept with the aim to demonstrate VLBI between a balloon-borne telescope and a ground-based telescope. We have presented the design of the 36 inch BVEX radio telescope and 367 K receiver which is a cheap and scalable alternative built from commercially available off the shelf components. We also discussed a novel sub-mm precision position tracking system built from readily available commercial-off-the-shelf components. While the 2025 flight from Timmins, Ontario was not successful due to a major leak in the launch balloon, we are upgrading the telescope for a second flight opportunity from Palmas, Tocantins, Brazil, in 2027 with an improved receiver temperature of 155 K. If BVEX is successful then future-generation high-frequency versions of BVEX could help improve the dynamic range of ground-based VLBI arrays such as the Event Horizon Telescope (EHT) as well as the Global mm-VLBI Array (GMVA).

\acknowledgments % equivalent to \section*{ACKNOWLEDGMENTS}       
The Balloon-borne VLBI Experiment (BVEX) was primarily funded by the Canadian Space Agency's (CSA) Flights and
Fieldwork for the Advancement of Science and Technology (FAST) 2021 AO (21FAQUEA18), with supporting funding from the Queen's Faculty of
Arts Infrastructure Fund,  the Canada Foundation for Innovation (CFI) John R. Evans
Leaders Fund (JELF Project Number 43766), National Science and Engineering Research Council
(NSERC)  Discovery Grant RGPIN/06266-2020, and
the Queen's University Research Initiation
Grant. We would also like to acknowledge observing and planning support from the Very Long Baseline Array (VLBA) of the National Radio Astronomy Observatory (NRAO), the Max-Planck Institut für Radioastronomie (MPIfR) and the Radio Telescope Effelsberg, and MIT's Haystack Observatory. Furthermore, we would like to thank Jens Kauffmann, Assistant Director of the MIT Haystack Observatory, and the Haystack Observatory staff for their support of our project in ground-based testing as well as simultaneous observation support. We would like to thank Frank Jiang as well as Doug Henke from the National Research Council of Canada (NRC)'s Herzberg Astronomy and Astrophysics Research Centre (HAA) for their advice during the receiver design process. Finally, we would like to thank the CSA STRATOS and CNES teams for supporting our 2025 Timmins flight campaign.

\appendix
\section{DERIVATION OF THE ACCELEROMETER DRIFT}
\label{app:derivation}
	We begin by modeling the behavior of the accelerometer by a random walk process as follows
	\begin{equation}
		v_n = \sum_{i=1}^{n} a_i \Delta t
	\end{equation}
	where $v_n$ is the change in velocity at time $t_n$ corresponding to the $n$-th sample since the start of the integration, $\Delta t$ is the time step between accelerometer samples and $a_i$ is the measured acceleration at time $t_i$ which for simplicity, we consider to be an independent random variable with $\mu$=0, $\sigma=\delta a$ where $\delta a$ is the accelerometer uncertainty. We can also do this for non-zero mean but the results for the variance will be the same. We now want to find the variance in $v_n$:
	\begin{equation}
		v_n^2 =\Delta t^2\left(\sum_{i=1}^{n} a_i^2 +2\sum_{1\leq i<j \leq n}a_i a_j\right).
	\end{equation}
    Because we (without loss of generality) assume that $a_i$ has zero mean $v_n$ will also have zero mean and the variance is given by
    \begin{equation}
    	Var(v_n)=E(v_n ^2) = \Delta t^2 \left(\sum_{i=1}^{n} E(a_i^2) +2\sum_{1\leq i<j \leq n}E(a_i a_j)\right)
    \end{equation}
	where $E(x)$ denotes the expectation value of random variable $x$. The latter term will be zero because $a_i$ is independent and therefore:
	\begin{equation}
		Var(v_n)=\delta v^2= \Delta t^2 n \delta a^2.
	\end{equation}
	We can put this into accelerometer time by noting that $\Delta t = 1/f_{smp}$ and $n=tf_{smp}$ where $f_{smp}$ is the accelerometer sampling rate. Therefore the expected error in the integrated velocity change at an arbitrary time t is:
	\begin{equation}
		\delta v =\delta a \sqrt{\frac{t}{f_{smp}}}
	\end{equation}

	Now we can consider the change in position which we model by:
	\begin{equation}
		\delta x_n = \sum_{i=1}^{n} v_i \Delta t.
	\end{equation}
	Just as before we get that
	\begin{equation}
		Var(x_n)=E(x_n ^2) = \Delta t^2 \left(\sum_{i=1}^{n} E(v_i^2) +2\sum_{1\leq i<j \leq n}E(v_i v_j)\right)
	\end{equation}
	however, in this case, the second term no longer vanishes because the velocity now is highly correlated. Therefore:
	\begin{equation}
		Var(x_n)= \Delta t^2 \left(n\delta v^2 +2\sum_{1\leq i<j \leq n}E\left(\sum_{k=1}^{i} a_k \Delta t \sum_{l=1}^{j} a_l \Delta t\right)\right),
	\end{equation}
	\begin{equation}
		Var(x_n)= \Delta t^4 \left(n^2\delta a^2 +2\sum_{1\leq i<j \leq n}E\left(\sum_{k=1}^{i} a_k  \sum_{l=1}^{j} a_l \right)\right).
	\end{equation}
	Because $a_i$ is independent, the cross-terms in the second term vanish when we take the expectation value, and since we require that $i<j$ we therefore get
	\begin{equation}
	 	Var(x_n)= \Delta t^4 \left(n^2\delta a^2 +2\sum_{1\leq i<j \leq n}E\left(\sum_{k=1}^{i} a_k^2\right)\right),
	\end{equation}
	\begin{equation}
		Var(x_n)= \Delta t^4 \left(n^2\delta a^2 +2\sum_{1\leq i<j \leq n}i\delta a^2\right).
	\end{equation}
	We can rewrite the remaining sum as
	\begin{equation}
		Var(x_n)= \Delta t^4 \left(n^2\delta a^2 +2\delta a^2\sum_{j=2}^{n}\sum_{i=1}^{j-1} i\right),
	\end{equation}
    using that $\sum_{i=1}^{n}i=\frac{n(n+1)}{2}$
    \begin{equation}
		Var(x_n)= \Delta t^4 \left(n^2\delta a^2 +2\delta a^2\sum_{j=2}^{n}\frac{(j-1)j}{2}\right),
	\end{equation}
	\begin{equation}
		Var(x_n)= \Delta t^4 \left(n^2\delta a^2 +\delta a^2 \left(\sum_{j=1}^{n}(j+1)j - (n+1)n\right)\right),
	\end{equation}
	Once again, using that $\sum_{i=1}^{n}i=\frac{n(n+1)}{2}$ and $\sum_{i=1}^{n}{i^2}=\frac{n(n+1)(2n+1)}{6}$ we get:
	\begin{equation}
		Var(x_n)= \Delta t^4 \left(n^2\delta a^2 +\delta a^2\left(
		\frac{n(n+1)(2n+1)}{6}-\frac{n(n+1)}{2}\right)\right),
	\end{equation}
	\begin{equation}
		Var(x_n)= \frac{\Delta t^4 \delta a^2}{3}\left(3n^2 +n^3-n\right).
	\end{equation}
	We can convert back to time just as we did before
	\begin{equation}
		Var(x_n)= \frac{\delta a^2}{3}\left(3\left(\frac{t}{f_{smp}}\right)^2 +\frac{t^3}{f_{smp}}-\frac{t}{f_{smp}^3}\right)
	\end{equation}
	and hence:
	\begin{equation}
		\delta x =\frac{\delta a}{\sqrt{3}}\sqrt{3\left(\frac{t}{f_{smp}}\right)^2 +\frac{t^3}{f_{smp}}-\frac{t}{f_{smp}^3}}.
	\end{equation}

% References
\bibliography{report} % bibliography data in report.bib
\bibliographystyle{spiebib} % makes bibtex use spiebib.bst

\end{document}